# Quantum computing in molecular magnets

Michael N. Leuenberger & Daniel Loss

*Department of Physics and Astronomy, University of Basel, Klingelbergstrasse 82, 4056 Basel, Switzerland*

**Shor and Grover demonstrated that a quantum computer can outperform any classical computer in factoring numbers[1] and in searching a database[2] by exploiting the parallelism of quantum mechanics. Whereas Shor's algorithm requires both superposition and entanglement of a many-particle system[3], the superposition of single-particle quantum states is sufficient for Grover's algorithm[4]. Recently, the latter has been successfully implemented[5] using Rydberg atoms. Here we propose an implementation of Grover's algorithm that uses molecular magnets[6, 7, 8, 9, 10], which are solid-state systems with a large spin; their spin eigenstates make them natural candidates for single-particle systems. We show theoretically that molecular magnets can be used to build dense and efficient memory devices based on the Grover algorithm. In particular, one single crystal can serve as a storage unit of a dynamic random access memory device. Fast electron spin resonance pulses can be used to decode and read out stored numbers of up to $10^5$, with access times as short as $10^{-10}$ seconds. We show that our proposal should be feasible using the molecular magnets $Fe_8$ and $Mn_{12}$.**

Suppose we want to find a phone number in a phone book consisting of $N=2^n$ entries. Usually it takes $N/2$ queries on average to be successful. Even if the $N$ entries were encoded binary, a classical computer would need approximately $\log_2 N$ queries to find the desired phone number[2]. But the computational parallelism provided by the superposition and interference of quantum states enables the Grover algorithm to reduce the search to one single query[2]. Here we will show that this query can be implemented in terms of a unitary transformation applied to the single spin of a molecular magnet. Such molecular magnets, forming identical and largely independent units, are embedded in a single crystal so that the ensemble nature of such a crystal provides a natural amplification of the magnetic moment of a single spin. However, for the Grover algorithm to succeed, it is necessary to find ways to generate arbitrary superpositions of spin eigenstates. For spins larger than 1/2 this turns out to be a highly non-trivial task as spin excitations induced by magnetic dipole transitions in conventional electron spin resonance (ESR) can occur only in discrete steps of one $\hbar$ (Planck's constant divided by $2\pi$), that is, single steps by two or more $\hbar$ values are excluded by selection rules. To circumvent such physical limitations we propose an unusual scenario which, in principle, allows the controlled generation of arbitrary spin superpositions through the use of multifrequency coherent magnetic radiation in the microwave and radiofrequency range. In particular, we will show by means of the *S*-matrix and time-dependent high-order perturbation theory that by using advanced ESR techniques it is possible to coherently populate and manipulate many spin states simultaneously by applying one single pulse of a magnetic a.c. field containing an appropriate number of matched frequencies. This a.c. field creates a nonlinear response of the magnet via multiphoton absorption processes involving particular sequences of $\sigma$ and $\pi$ photons which allows the encoding and, similarly, the decoding of states. Finally, the subsequent read-out of the decoded quantum state can be achieved by means of pulsed ESR techniques. These exploit the non-equidistance of energy levels which is typical of molecular magnets. The method we propose here is interesting in its own right since there has never been an experimental or theoretical attempt, to our knowledge, that shows that the states of spin systems with $s > 1/2$ can be coherently populated.

We implement the Grover algorithm in the single-spin representation with the level spectrum of a spin system as shown in Fig. 1. First, a strong magnetic field in *z* direction must be applied in order to prepare the initial state $|\psi_0\rangle = |s\rangle$. Then this field is reduced almost to zero (up to the bias $\delta H_z$) in such a way that all $|m\rangle$-states are localized, say, on the left side of the potential barrier. Thus, the magnetic moment pointing along the $z$ axis assumes its maximum value and the single spin of a molecular magnet is described by the hamiltonian $H_{\text{spin}} = H_a + V$ (see Fig. 1). To mark specific states, with a certain occupation amplitude (including phases), we could apply weak oscillating transverse magnetic fields $\mathbf{H}_\perp$ to induce multiphoton transitions via virtual states, which can usually be calculated in perturbation theory in $\mathbf{H}_\perp$. However, the Grover algorithm requires that all the $k$-photon transitions, $k = 1,2,...,s-1$, have (approximately) the same amplitudes (and possibly different phases). Thus, this would require that all terms of power $V^1$, $V^2$, ... , $V^{s-1}$ must be of comparable magnitude. Obviously, perturbation theory breaks down in such a case.

To bypass this problem, in our method, all the transition amplitudes between the states $|s\rangle$ and $|m\rangle$, $m=1,2,..., s-1$, are of the same order in perturbation $V$. This allows us to use perturbation theory. It works only if the energy levels are not equidistant, which is typically the case in molecular magnets owing to anisotropies (in contrast to, for example, a harmonic oscillator potential). In general, if we choose to work with the states $m = m_0, m_0+1,...,s-1$, where $m_0 = 1,2,...,s-1$, we have to go up to $n$th order in perturbation, where $n = s-m_0$ is the number of computational states used for the Grover search algorithm (see below), to obtain the first non-vanishing contribution. Figure 2 shows the transitions for $s = 10$ and $m_0 = 5$. The $n$th-order transitions correspond to the nonlinear response of the spin system to strong magnetic fields. Thus, a coherent magnetic pulse of duration $T$ is needed with a discrete frequency spectrum $\{\omega_m\}$, say, for $Mn_{12}$ between 20 and 300 GHz and a single low-frequency $\omega_0$ around 100 MHz (for pulse shaping techniques see ref. 11 and references therein). The low-frequency field $\mathbf{H}_z(t) = H_0(t)\cos(\omega_0 t)\mathbf{e}_z$, applied along the easy-axis, couples to the spin of the molecular magnet through the hamiltonian

$$V_{\text{low}}(t) = g\mu_B H_0(t)\cos(\omega_0 t)S_z \qquad (1)$$

where $\hbar\omega_0 \ll e_{m_0} - e_{m_0+1}$ and $\mathbf{e}_z$ is the unit vector pointing along the $z$ axis. The $\pi$ photons[12] of $V_{\text{low}}$ supply the necessary energy for the resonance condition (see below). They give rise to virtual transitions with $\Delta m = 0$, that is, they do not transfer any angular momentum, see Fig. 2.

The perturbation hamiltonian for the high-frequency transitions from $|s\rangle$ to virtual states that are just below $|m\rangle$, $m = m_0,\ldots,s-1$, given by the transverse fields $\mathbf{H}_\perp^-(t) = \sum_{m=m_0}^{s-1} H_m(t)\left[\cos(\omega_m t + \Phi_m)\mathbf{e_x} - \sin(\omega_m t + \Phi_m)\mathbf{e}_y\right]$, reads

$$\begin{aligned}V_{\text{high}}(t) &= \sum_{m=m_0}^{s-1} g\mu_B H_m(t)\left[\cos(\omega_m t + \Phi_m)S_x - \sin(\omega_m t + \Phi_m)S_y\right] \\ &= \sum_{m=m_0}^{s-1} \frac{g\mu_B H_m(t)}{2}\left[e^{i(\omega_m t + \Phi_m)}S_+ + e^{-i(\omega_m t + \Phi_m)}S_-\right]\end{aligned} \qquad (2)$$

with phases $\Phi_m$ (see below), where we have introduced the unit vectors $\mathbf{e}_x$ and $\mathbf{e}_y$ pointing along the $x$ and $y$ axis, respectively. These transverse fields rotate clockwise and thus produce left circularly polarized $\sigma^-$ photons which induce only transitions in the left well (see Fig. 1). In general, absorption (emission) of $\sigma^-$ photons gives rise to $\Delta m = -1$ ($\Delta m = +1$) transitions, and vice versa in the case of $\sigma^+$ photons. Anti-clockwise rotating magnetic fields of the form $\mathbf{H}_\perp^+(t) = \sum_{m=m_0}^{s-1} H_m(t)\left[\cos(\omega_m t + \Phi_m)\mathbf{e_x} + \sin(\omega_m t + \Phi_m)\mathbf{e}_y\right]$ can be used to induce spin transitions only in the right well (see Fig. 1). In this way, both wells can be accessed independently.

Next we calculate the quantum amplitudes for the transitions induced by the magnetic a.c. fields (see Fig. 2) by evaluating the S-matrix perturbatively. The $j$th-order term of the perturbation series of the S-matrix in powers of the total perturbation hamiltonian $V(t) = V_{\text{low}}(t)+V_{\text{high}}(t)$ is expressed by

$$S_{m,s}^{(j)} = \left(\frac{1}{i\hbar}\right)^j \prod_{k=1}^{j-1} \int_{-\infty}^{\infty} dt_k \int_{-\infty}^{\infty} dt_j \Theta(t_k - t_{k+1}) U(\infty,t_1)V(t_1)U(t_1,t_2)V(t_2)\ldots V(t_j)U(t_j,-\infty), \quad (3)$$

which corresponds to the sum over all Feynman diagrams of order $j$, and where $U(t,t_0) = e^{-iH(t-t_0)/\hbar}$ is the free propagator, $\Theta(t)$ is the Heavyside function. The total S-matrix is then given by $S = \sum_{j=0}^{\infty} S^{(j)}$. The high-frequency virtual transition changing $m$ from $s$ to $s$-1 is induced by the frequency $\omega_{s-1} = \omega_{s-1,s}-(n-1)\omega_0$. The other high frequencies $\omega_m$, $m = m_0,\ldots,s$-2, of the high-frequency fields $H_m$ mismatch the level separations by $\omega_0$, that is, $\hbar\omega_m = \varepsilon_m - \varepsilon_{m+1} + \hbar\omega_0$, see Fig. 2. As the levels are not equidistant, it is possible to choose the low and high frequencies in such a way that $S_{m,s}^{(j)} = 0$ for $j < n$, in which case the resonance condition is not satisfied, that is, energy is not conserved. In addition, the higher-order amplitudes $\left|S_{m,s}^{(j)}\right|$ are negligible compared to $\left|S_{m,s}^{(n)}\right|$ for $j > n$. Using rectangular pulse shapes, $H_k(t) = H_k$, if $-T/2 < t < T/2$, and 0 otherwise, for $k = 0$ and $k \geq m_0$, we obtain after lengthy but straightforward calculation ($m \geq m_0$)

$$S_{m,s}^{(n)} = \sum_F \Omega_m \frac{2p}{i}\left(\frac{g\mathbf{m}_B}{2\hbar}\right)^n \frac{\prod_{k=m}^{s-1} H_k e^{i\Phi_k} H_0^{m-m_0} p_{m,s}(F)}{(-1)^{q_F} q_F! r_s(F)! \mathbf{w}_0^{n-1}}$$
$$\times \mathbf{d}^{(T)}\left(\mathbf{w}_{m,s} - \sum_{k=m}^{s-1} \mathbf{w}_k - (m-m_0)\mathbf{w}_0\right) \quad (4)$$

where $\Omega_m = (m-m_0)!$ is the symmetry factor of the Feynman diagrams $F$ (see Fig. 2), $q_F = m-m_0-r_s(F)$, $p_{m,s}(F) = \prod_{k=m}^{s} \langle k|S_z|k\rangle^{r_k(F)} \prod_{k=m}^{s-1} \langle k|S_-|k+1\rangle$, $r_k(F) = 0,1,2,\ldots \leq m-m_0$ is the number of $\pi$ transitions directly above or below the state $|k\rangle$, depending on the particular Feynman diagram $F$, and $\mathbf{d}^{(T)}(\mathbf{w}) = \frac{1}{2p}\int_{-T/2}^{+T/2} e^{i\mathbf{w}t} dt = \sin(\mathbf{w}T/2)/p\mathbf{w}$ is the delta-function of width $1/T$, ensuring overall energy conservation (resonance condition) for $\omega T \gg 1$. The duration $T$ of the magnetic pulses must be shorter than the lifetimes $\tau_d$ of the states $|m\rangle$ (see Fig. 1). For illustration, we now focus on the case $s = 10$, $m_0 = 5$, and thus $n = 5$, described by $S_{m,10}^{(5)}$, since this is most relevant for the molecular magnets $Mn_{12}$ and $Fe_8$ (see Fig. 2). In general, the Grover algorithm requires that the levels are simultaneously populated with roughly equal amplitudes, that is, $\left|S_{m,s}^{(n)}\right| \approx \left|S_{s-1,s}^{(n)}\right|$, $\forall m \geq m_0$, from which we can deduce the required field amplitudes using equation (4)

$$|H_8/H_0|=0.04,\ |H_7/H_0|=0.25,\ |H_6/H_0|=0.61,\ |H_5/H_0|=1.12 \qquad (5)$$

This means that the fields $H_0$ and $H_9$, and the frequency $\omega_0$ can be chosen independently. We note particularly that the amplitudes $H_k$ do not differ too much from each other, which can be traced back to partial cancellations in equation (4) owing to the factor $(-1)^{q_F}$. This fact is most useful for applications.

Next, we estimate the transition rate needed to coherently populate the five levels with one single pulse with $V_{\text{low}} + V_{\text{high}}$. As $[\delta^{(T)}(\omega)]^2 \approx (T/2\pi)\delta^{(T)}(\omega)$, the transition rate $w_{m,s} = \left|S_{m,s}^{(n)}\right|^2 / T$ from $|10\rangle$ to $|5\rangle$ at resonance (with $\delta^{(T)}(0) = T/2\pi$) reads

$$w_{5,10} = T\left(\frac{g m_B}{2\hbar}\right)^{10} \left|\frac{H_5 H_6 H_7 H_8 H_9 p_{5,10}}{4!\, w_0^4}\right|^2, \qquad (6)$$

where $p_{5,10} = \prod_{k=5}^{9} \langle k|S_-|k+1\rangle$. We then insert the parameters $\omega_0 = 5\times10^8$ s$^{-1}$, $T = 10^{-7}$ s, $H_0 = H_9 = 20$ G giving the transition rate $w_{5,10} = 9\times10^6$ s$^{-1}$, which is of the order of $\tau_d$, that is, $Tw_{5,10} = 1$. For our purpose it is sufficient to choose $Tw_{5,10} \ll 1$. If $\omega_0 = 5\times10^7$ s$^{-1}$, $T = 10^{-7}$ s, $H_0 = H_9 = 2$ G, we obtain the transition rate $w_{5,10} = 9\times10^4$ s$^{-1}$, thus $Tw_{5,10} \approx 0.01$. Thus we have shown that the required amplitudes and frequencies of the fields given in equation (1) and (2) are experimentally accessible.

We adapt now the Grover scheme[5] to describe the quantum computational read-in and decoding of the quantum data register $a$. For simplicity we set the relative phase $\Phi_0 = 0$ of the low-frequency field $\mathbf{H}_z(t)$; see equation (1).

(1) Read-in. We start from the ground state $|s\rangle$ as initial state. Then, in order to introduce the desired phases $\Phi_m$ for each state $|m\rangle$ we need to irradiate the system with a coherent magnetic pulse of duration $T$ containing the $n$ high-frequency fields $H_m[\cos(\omega_m t+\Phi_m)e_x - \sin(\omega_m t+\Phi_m)e_y]$ (see equation (2)) with $\Phi_m = \sum_{k=s-1}^{m+1} \Phi_k + j_m$ ($\varphi_m$ are the relative phases), and the low-frequency field $H_z(t)$ (see equation (1)), yielding $S_{m_0,s}^{(n)} = \ldots = S_{s-2,s}^{(n)} = S_{s-1,s}^{(n)} = \pm h$, $\eta > 0$, that is, $\varphi_m = 0, \pi$, depending on the number to be encoded. For example, to encode the number $13_{10} = 1101_2$ in Fig. 2, we need $\varphi_9 = \varphi_8 = \varphi_7 = 0$ and $\varphi_6 = \varphi_5 = \pi$, where the states $m = 9, 8, 7, 6, 5$ represent the binary digits $2^0, 2^1, 2^2, 2^3, 2^4$, respectively. We note that $S_{m,s}^{(n)}$ can be either positive or negative, depending on the explicit Feynman diagrams $F$. In this way one can prepare the quantum data register $a = (a_s, a_{s-1}, \ldots, a_{m_0})$, where the bits $a_s = 1$, $a_m = \pm \eta$ are the amplitudes of the state $|y\rangle = \sum_{m=m_0}^{s} a_m |m\rangle$. This pulse performs a unitary transformation up to order $\eta$, that is, $U_{m,s} + O(h^2) = S_{m,s}^{(n)}$. In this way an arbitrary integer between 0 and $2^n$ can be stored.

(2) Decoding. In order to decode the phase information stored in the data register $a$, a universal single pulse (see equations (1) and (2)) leading to $S_{m_0,s}^{(n)} = \ldots = S_{s-2,s}^{(n)} = S_{s-1,s}^{(n)} = -h$ must be applied, which performs approximately a unitary transformation for $\eta \ll 1$. The accumulated error is about $n\eta^2$, which must be kept smaller than 1. Because $\eta > \eta_0 > 0$, with $\eta_0$ given by the precision of the detection, we require that $n \ll 1/h_0^2$. We note that this decoding works also if the bits $a_m$, with $\eta_0 < |a_m| \ll 1$, have different amplitudes (but still of similar magnitude). This pulse amplifies the flipped bits, which have amplitude $-\eta$, and suppresses the rest of the bits with amplitude $+\eta$. For the example shown in Fig. 2 the relative phases in equation (2) must be set $\varphi_9 = \varphi_7 = \varphi_5 = 0$ and $\varphi_8 = \varphi_6 = \pi$ (irrespective of $a$).

(3) Read-out. Once a state has been marked and amplified, that is, decoded, we must be able to read out this information physically. This task can be accomplished by standard spectroscopy with, say, pulsed ESR, where the circularly polarized radiation can now be incoherent because we need the absorption intensity of only one pulse. Full spectral analysis of $Mn_{12}$ have been performed with ESR[13] and neutron scattering[14]. Thus, we can assume that the spectrum is known. Now, irradiation of the magnet with a single pulse of duration $T$ containing the frequencies $\omega_{m-1,m}$, $m = s-2, \ldots, m_0$, induces transitions that are described by the first-order amplitudes $S_{m-1,m}^{(1)}$ (higher-order multiphoton effects can now be neglected for sufficiently small fields). For instance, let us assume that the state $|7\rangle$ is marked, that is, populated. Then, we would observe stimulated emission for the transitions from $|7\rangle$ to $|8\rangle$ at the frequency $\omega = \omega_{7,8}$ and stimulated absorption of approximately the same intensity for the transition from $|7\rangle$ to $|6\rangle$ at $\omega = \omega_{6,7}$, which uniquely identifies the marked level, because the levels are not equidistant. Generally, if the states $|m_1\rangle, |m_2\rangle, \ldots, |m_k\rangle$, where $1 \leq k \leq n$, are marked, the following absorption/emission intensity in leading order is measured:

$$I_{s-2}^{m_0} = \sum_{i=1}^{k} \left( \left| S_{m_i-1,m_i}^{(1)} \right|^2 + \left| S_{m_i+1,m_i}^{(1)} \right|^2 \right). \tag{7}$$

This spectrum identifies all the marked states unambiguously. We emphasize that the entire Grover algorithm (read-in, decoding, read-out) requires three subsequent pulses each of duration $T$ with $t_d > T > w_0^{-1} > w_m^{-1}, w_{m,m\pm1}^{-1}$. This gives a 'clock-speed' of about 10 GHz for $Mn_{12}$, that is, the entire process of read-in, decoding, and read-out can be performed within about $10^{-10}$ s.

Finally, so far we have used only the left well of the potential in Fig. 1. As both wells can be accessed separately during read-in, decoding and read-out by magnetic fields that rotate either clockwise or anti-clockwise, a single molecular magnet represents a two-digit number. Every digit can store $N = 2^{s-1}$ elements at most, which allows us to store a number between 0 and $2^{2s-2} = 2.6 \times 10^5$ in a single crystal made of molecular magnets with spin $s = 10$. If $M > 2$ phases can be distinguished for $\varphi_m$ (depending on the experimental resolution)[5], a number between 0 and $M^{2s-2}$ can be stored in a single crystal of molecular magnets with spin $s$. We note that the experimental overhead required by the Grover search algorithm involves only the control of $\log_M N$ frequencies, which, once available, can decode any number between 1 and $N$ by means of a single magnetic pulse. Our proposal for implementing Grover's algorithm works not only for molecular magnets but for any electron or nuclear spin system with non-equidistant energy levels. Although such spin systems cannot be scaled to arbitrarily large spin *s* — the larger a spin becomes, the faster it decoheres and the more classical its behavior will be — we can use such spin systems of given *s* to great advantage in building dense and highly efficient memory devices.

**Acknowledgements**

We thank G. Salis and J. Schliemann for useful comments. This work has been supported in part by the Swiss NSF and by the European Union Molnanomag network.

Correspondence should be addressed to D. L. (e-mail: Daniel.Loss@unibas.ch).

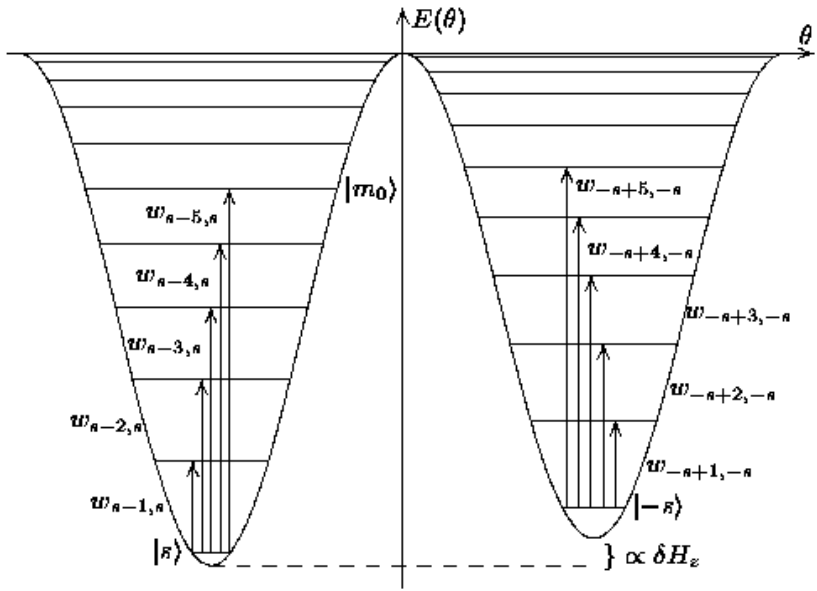

**Figure 1** Double well potential seen by the spin due to magnetic anisotropies in $Mn_{12}$. Arrows depict transitions between spin eigenstates driven by the external magnetic field **H** (see below). Molecular magnets have the important advantage that they can be grown naturally as single crystals of up to 10–100 μm length containing about $10^{12}$ to $10^{15}$ (largely) independent units so that only minimal sample preparation is required. The molecular magnets behave like single spins, for example, $Mn_{12}$ (refs 7, 8) and $Fe_8$ (refs 9, 10) have a spin $s = 10$ ground state. Thus, they can be described by a single-spin hamiltonian of the form $H_{spin} = H_a + V + H_{sp} + H_T$[15,16], where $H_a = -AS_z^2 - BS_z^4$ represents the magnetic anisotropy ($A >> B > 0$), that is, the easy axis of the spin lies along the $z$ direction. The Zeeman term $V = g\mu_B \mathbf{H} \cdot \mathbf{S}$ describes the coupling between the external magnetic field **H** and the spin **S** of length $s$. The calculational states are given by the $2s+1$ eigenstates $|m\rangle$ of $H_a + g\mu_B \delta H_z S_z$ with eigenenergies $\varepsilon_m = -Am^2 - Bm^4 + g\mu_B \delta H_z m$, $-s \leq m \leq s$. The corresponding classical anisotropy potential energy $E(\theta) = -As\cos^2\theta - Bs\cos^4\theta + g\mu_B \delta H_z s \cos\theta$, shown here, is obtained by the substitution $S_z = s\cos\theta$, where $\theta$ is the polar spherical angle. We have introduced the notation $\hbar \omega_{m,m'} = \varepsilon_{m'} - \varepsilon_m$. The hamiltonian $H_T$ induces tunnelling between (quasi) degenerate $|m\rangle$-states with tunnel splitting energy $E_{mm'}$. However, applying a bias field $\delta H_z$ such that $g\mu_B \delta H_z > E_{mm'}$, tunnelling can be completely suppressed and thus $H_T$ can be neglected[15,16]. We also assume temperatures of below 1 K such that transitions due to spin-

phonon interactions ($H_{sp}$) can also be neglected. In this regime, the level lifetime in Fe$_8$ and Mn$_{12}$ is estimated to be about $\tau_d = 10^{-7}$s, limited mainly by hyperfine and/or dipolar interactions[10,15,16].

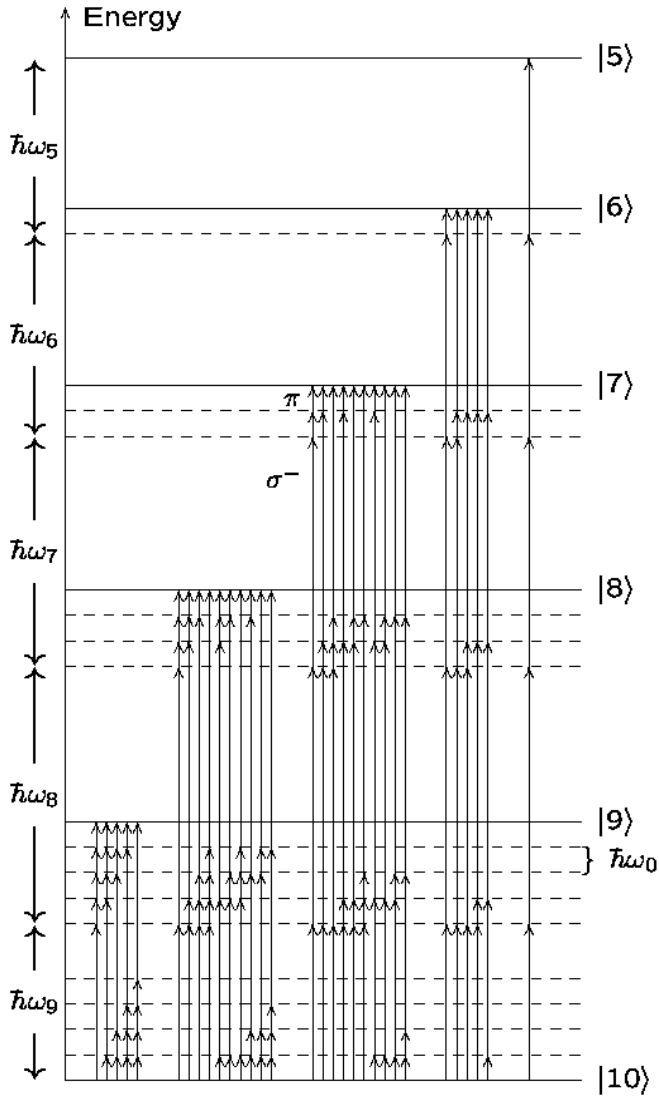

**Figure 2** Feynman diagrams $F$ that contribute to $S_{m,s}^{(5)}$ for $s=10$ and $m_0=5$ describing transitions (of 5th order in $V$) in the left well of the spin system (see Fig. 1). The blue (red) arrows indicate the transitions induced by the high-(low-)frequency magnetic fields $H_m$ ($H_0$) shown in equation (2) (equation (1)). Whereas the $H_m$ fields transfer angular momentum to the spin of the molecular magnet by means of $\sigma^-$ photons, the $H_0$ field provides only energy without angular momentum by means of $\pi$ photons. In this way all the transition amplitudes $S_{m,s}^{(5)}$ are of similar

magnitude (see text). We note that $S_{m,s}^{(j)} = 0$ for $j < n$, and $\left|S_{m,s}^{(j)}\right| << \left|S_{m,s}^{(n)}\right|$ for $j > n$. For example, the transition from $|10\rangle$ to $|7\rangle$ arises from the absorption of five photons in total, comprising three $\sigma^-$ photons and two $\pi$ photons. Since it does not matter in which order the five photons are absorbed, there are $\binom{5}{2} = 10$ different kinds of diagrams. Also, it is essential that the levels are not equidistant; if they are, this scheme for multiphoton absorption does not work, because then resonances can occur already in lower order of the *S*-matrix leading to incoherent populations of the levels. There is a global phase factor due to the unperturbed time evolution of the spin system, given by $e^{-i\mathbf{e}_m(t-t_0)/\hbar}$, which can be easily accounted for and thus shall be ignored here. For a first test of the nonlinear response described here, we can irradiate the molecular magnet with an a.c. field of frequency $\omega_{s-1,s}/2$, which gives rise to a two-photon absorption and thus to a Rabi oscillation between the states $|s\rangle$ and $|s-1\rangle$. We note that for stronger magnetic fields it is in principle possible to generate superpositions of Rabi oscillations between the states $|s\rangle$ and $|s-1\rangle$, $|s\rangle$ and $|s-2\rangle$, $|s\rangle$ and $|s-3\rangle$, and so on (to be published elsewhere).